\documentclass[conference]{IEEEtran}
\IEEEoverridecommandlockouts
\usepackage{cite}
\usepackage{amsmath,amssymb,amsfonts}
\usepackage{algorithmic}
\usepackage{graphicx}
\usepackage{textcomp}
\usepackage{xcolor}
\usepackage{comment}

\def\BibTeX{{\rm B\kern-.05em{\sc i\kern-.025em b}\kern-.08em
    T\kern-.1667em\lower.7ex\hbox{E}\kern-.125emX}}

\usepackage{subcaption}

\usepackage[acronym,shortcuts]{glossaries}
\newcommand\norm[1]{\left\lVert#1\right\rVert}
\newcommand{\rom}[1]{\uppercase\expandafter{\romannumeral #1\relax}}

\newacronym{b5g}{B5G}{beyond fifth generation}
\newacronym{mmWave}{mmWave}{millimeter wave}
\newacronym{us}{US}{uncorrelated scattering}
\newacronym{wss}{WSS}{wide-sense stationarity}
\newacronym{dpss}{DPSS}{discrete prolate spheroidal sequence}
\newacronym{dsft}{DSFT}{discrete symplectic Fourier transform}
\newacronym{lsf}{LSF}{local scattering function}
\newacronym{v2v}{V2V}{vehicle-to-vehicle}
\newacronym{suv}{SUV}{sport utility vehicle}
\newacronym{los}{LOS}{line-of-sight}
\newacronym{nlos}{NLOS}{non-LOS} 
\newacronym{rms}{RMS}{root mean square}

\newacronym[firstplural={multipath components
(MPCs)}]{mpc}{MPC}{multipath component}

\begin{document}

\title{Evaluation of stationarity regions in measured non-WSSUS 60\,GHz mmWave V2V channels\\
}

\author{\IEEEauthorblockN{Danilo Radovic}
\IEEEauthorblockA{\textit{Institute of Telecommunication} \\
\textit{TU Wien}\\
Vienna, Austria \\
danilo.radovic@tuwien.ac.at}
\and
\IEEEauthorblockN{ Herbert Groll}
\IEEEauthorblockA{\textit{Institute of Telecommunication} \\
\textit{TU Wien}\\
Vienna, Austria \\
 herbert.groll@tuwien.ac.at}
\and
\IEEEauthorblockN{Christoph F. Mecklenbräuker}
\IEEEauthorblockA{\textit{Institute of Telecommunication} \\
\textit{TU Wien}\\
Vienna, Austria \\
}
}

\maketitle

\begin{abstract}
Due to high mobility in multipath propagation environments, vehicle-to-vehicle (V2V) channels are generally time and frequency variant. Therefore, the criteria for wide-sense stationarity (WSS) and uncorrelated scattering (US) are just satisfied over very limited intervals in the time and frequency domains, respectively. We test the validity of these criteria in measured vehicular $60$ GHz millimeter wave (mmWave) channels, by estimating the local scattering functions (LSFs) from the measured data.
Based on the variation of the LSFs, we define time-frequency stationarity regions, over which the WSSUS assumption is assumed to be fulfilled approximately.  We analyze and compare both line-of-sight (LOS) and non-LOS (NLOS) V2V communication conditions. 

\noindent
We observe large stationarity regions for channels with a dominant LOS connection, without relative movement between the transmitting and receiving vehicle. 

In the same measured urban driving scenario, modified by eliminating the LOS component in the post-processing, the channel is dominated by specular components reflected from an overpassing vehicle with a relative velocity of $56$\,km/h. Here, we observe a stationarity bandwidth of $270$\,MHz. Furthermore, the NLOS channel, dominated by a single strong specular component, shows a relatively large average stationarity time of $16$\,ms, while the stationarity time for the channel with a rich multipath profile is much shorter, in the order of $5$\,ms.

\end{abstract}


\begin{figure}[b]
    \noindent\fbox{%
        \parbox{\dimexpr\linewidth-2\fboxsep-2\fboxrule\relax}{%
            © 2022 IEEE.  Personal use of this material is permitted. Permission from IEEE must be obtained for all other uses, in any current or future media, including reprinting/republishing this material for advertising or promotional purposes, creating new collective works, for resale or redistribution to servers or lists, or reuse of any copyrighted component of this work in other works.
        }%
    }
\end{figure}

\section{Introduction}
\Ac{b5g} wireless communication technology is supposed to enhance the current systems by offering new wide-bandwidth communication channels. Furthermore, new technologies are developed to suit ever extending requirements of vehicular wireless communication. As a suitable solution for this task, communication over \ac{mmWave} frequency band is proposed. However, the vehicular channels show challenging characteristics due to high mobility and rapidly changing scattering environment. Moreover, the Doppler spectrum may vary over the time-frequency domain, limiting the validity assumption of \ac{wss} \cite{b1}. Nevertheless, the changing channel environment causes variations in the delay spectrum, violating the \ac{us} criterion. Furthermore, as the carrier frequency increases, the Doppler shift becomes more severe. Hence, the before mentioned stationarity issues are magnified, and we conclude that \ac{mmWave} \ac{v2v} channels are in general non-\ac{wss}\ac{us}. \\
However, the validity of many channel models and design of wireless transceivers is dependent on \ac{wss}\ac{us} assumption. Therefore, it is important to analyze the size of stationarity regions, as time-frequency area, in which \ac{wss}\ac{us} criteria are approximately satisfied. \\
A theoretical approach to defining stationarity regions is given in \cite{b2}. Further, multiple papers investigate the non-\ac{wss}\ac{us} behavior of the measured channels. The authors in \cite{b9} and \cite{b10} show spatial variation phenomenon, observing the fading process and defining channel stationarity as a function of a distance from the original position. \cite{b10} analyses the effect of having a \ac{los} as a contrast to a \ac{nlos} connection, showing that the spatial variation is not as severe in \ac{los} as in the \ac{nlos} conditions. Experimental contributions for \ac{v2v} communication, to the identification of stationarity regions in time and frequency, for $5$\,GHz band, are presented in \cite{b3}. The authors in \cite{b8} analyze the stationarity in the time domain depending on different \ac{v2v} $5$\,GHz measurement scenarios. However, to the best of our knowledge, the stationarity investigation for $60$\,GHz band has not been shown yet for vehicular communication.\\
We analyze the behavior of a real measured \ac{v2v} $60$\,GHz channel for typical \ac{los} urban scenarios. Furthermore, in order to compare the results with a \ac{nlos} scenario, we modify the measured channel, by eliminating the \ac{los} component in the post-processing. We define \ac{lsf}, by using the concepts described in \cite{b2}, and follow its variation over time and frequency. By calculating collinearity between \ac{lsf}s, and setting a threshold, we define time and frequency, over which \ac{lsf} is approximately constant. Hence, we quantify the consecutive time-frequency regions with approximately satisfied \ac{wss}\ac{us} condition, called stationarity regions.\\
In Section \ref{Sec:lsf} we introduce a definition of \ac{lsf}, and discuss its parametrization. The measured \ac{v2v} \ac{mmWave} channel scenarios, analyzed in this paper, are described in Section \ref{sec_measScen}. In Section \ref{sec_statCr}, we introduce the criteria for defining the channel as stationary. Furthermore, in Section \ref{sec_StatReg}, we present the evaluation of stationarity regions for the defined scenarios.  Conclusions are presented in Section \ref{sec_Conclusion}.

\section{Local Scattering Function and Parameterization} \label{Sec:lsf}
The time-varying channels are characterized by the discrete channel transfer function, sampled with $T_s$ resolution is time and $f_s$ in frequency, written as
\begin{equation} \label{ch_tf}
    \boldsymbol{H}[s,q] = H(s T_s , q f_s).
\end{equation}
We define $s\in[1,\cdots,S]$ and $q \in [1,\cdots, Q]$, $s$ and $q$ being indices in time and frequency domain. Now, we can express the total measurement bandwidth and recorded time as $B = Q f_s $ and $T_{\text{total}} = S T_s$. Statistical properties of vehicular channels in general do not remain constant over an arbitrary time nor frequency. Therefore, we define time-frequency regions, consisting of $N$ samples in the time and $M$ in frequency domain, in which the stationarity requirements are satisfied. By sequencing the channel transfer function, given in (\ref{ch_tf}), in the stationarity regions we define the local transfer functions, $\boldsymbol{\hat{H}_{k_t,k_f}} [s',q']$. $s'\in [1,\cdots,N]$ and $q'\in [1,\cdots,M]$ denote local time and frequency indices, and $k_t$ and $k_f$ the indices of each local region in time and frequency
\begin{subequations}
\begin{eqnarray}
 k_t \in [1, \cdots, K_t], \hspace{5pt} K_t = \lfloor \dfrac{S-N}{\Delta_t} \rfloor  + 1,~\text{and} \\
 k_f \in [1, \cdots, K_f], \hspace{5pt} K_f = \lfloor \dfrac{Q-M}{\Delta_f} \rfloor  + 1.
\end{eqnarray}
\end{subequations}
$\Delta_t$ and $\Delta_f$ describe time and frequency shift between each two consecutive local transfer functions, expressed in number of samples. The operator $\lfloor \cdot \rfloor$ rounds the number to the nearest lower integer. We denote $\boldsymbol{\hat{H}_{k_t,k_f}}$ as a matrix form of $k_t^{\text{th}},k_f^{\text{th}}$ local transfer function, containing $N\times M$ elements of time and frequency samples. We have to consider that by taking larger stationarity region we risk the violation of \ac{wss}\ac{us} assumption, but gain higher \ac{lsf} delay-Doppler resolution.\\
In order to decrease the variance, we calculate multiple independent spectral estimates by applying multitaper estimator. Here, we define orthogonal data tapers, with each taper providing good protection against leakage, similarly as given in \cite{b3}. For tapering function, we use an index limited sequence, with energy concentrated within the selected bandwidth, known as \ac{dpss}. The total number of tapers is given by $IJ$, $I$ being the number of sequences in time and $J$ in frequency. When setting the number of tapers, we have to pay attention to the trade-off between the variance reduction and biasing increase. More precisely, by increasing the number of used tapers the variance decreases but at the cost of biasing enlargement \cite{b4}. When we increase the number of tapers, we also have to consider the appropriate energy concentration bandwidth $2W_t$, defined as a multiple of fundamental frequency 
\begin{equation}
    W_t = \dfrac{a}{N T_s }, \hspace{5pt} a> 1.
\end{equation}
Setting $W_t$ higher allows for the larger number of tapers with good leakage properties \cite{b4}. In the further text, we will express the bandwidth by providing the value of $N W_t  = \tfrac{a}{T_s}$. Here, we want to emphasize, by using \ac{dpss} with index limited maximal energy concentration in time, and band limited in the Doppler domain, increasing $W_t$ leads to decrease in the resolution in terms of Doppler. Analogously, we define the \ac{dpss} in frequency domain. Furthermore, we define the time-frequency taper function
\begin{equation} \label{tap_func}
    \boldsymbol{G_w}[s', q'] = \boldsymbol{u_i}[s'] \boldsymbol{\tilde{u}_j}[q'], \hspace{5pt} s' \in [1,\cdots, N], q' \in [1,\cdots,M],
\end{equation}
with $i \in [1,\cdots,I]$, $j \in [1,\cdots,J]$ and $w = i J +j$. We write the taper function (\ref{tap_func}) as a matrix $\boldsymbol{G_w}$, of dimension $N \times M$. Now we define the matrix form of windowed channel transfer function

	\begin{equation}
	    \boldsymbol{\hat{\mathcal{H}}^{(G_w)}_{k_t,k_f}} = \boldsymbol{\hat{H}_{k_t,k_f}} \odot \boldsymbol{G_w},
	\end{equation}
	where $\odot$ denotes the element-wise Hadamard product. Applying \ac{dsft}, we obtain tapered Doppler-variant impulse response 
	\begin{equation}
	    \boldsymbol{\hat{S}_{k_t,k_f}^{(G_w)}} = \boldsymbol{F}_N \boldsymbol{\hat{\mathcal{H}}^{(G_w)}_{k_t,k_f}} \boldsymbol{F}_M^H,
	\end{equation}
	$\boldsymbol{F}_i$ and $\boldsymbol{F}_i^H$ representing discrete Fourier transform (DFT) and inverse DFT (IDFT) matrix of size $i$. By applying $IJ$ orthogonal tapers, we create multiple realizations of the same channel, and we calculate the multitaper estimate of the \ac{lsf} with uniform weighting across the tapered Doppler-variant impulse responses
	\begin{equation}
	    \boldsymbol{\hat{C}_{k_t,k_f}} = \dfrac{1}{IJ} \sum_{w=1}^{IJ} \boldsymbol{\hat{S}_{k_t,k_f}^{(G_w)}}.
	\end{equation}

\section{Measurement Scenarios} \label{sec_measScen}
\begin{figure}[ht]
    \centering
    \includegraphics[width=0.48\textwidth,page=3, trim=1.7cm 18.7cm 11cm 6cm, clip]{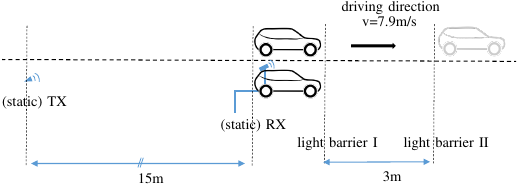}
    	\caption{Sketch of the measurement scenario.}
		\label{fig:sketchSc1}
\end{figure}

This work deals with the definitions of stationarity regions in measured \ac{v2v} \ac{mmWave} channels. The measurement campaign took place in an urban street environment, downtown Vienna, Austria. The channel is obtained at the central frequency $f_c = 60$\,GHz,  with a frequency resolution $f_s=4.96$\,MHz, and the $Q=103$ samples in frequency domain. Furthermore, the time resolution is $T_s=129.1$\,$\mu$s, and we are dealing with $S=5920$ time snapshots. The detailed description of measurement set-up may be found in \cite{b6}. \\
The transmitter and receiver are positioned on the right lane of an urban street, at fixed positions, on a tripod and a car's rear window, respectively. During the measurement, further vehicles are passing by on the left lane, see Fig. \ref{fig:sketchSc1}. Hence,  this scenario resembles an urban overtaking process.\\
At the transmitter side, a directive horn antenna is employed, directed along the street in driving direction. Hence, it covers the receiver and the overtaking vehicle within the $3$\,dB beam width of the antenna. Further, at the receiver, an open-ended waveguide antenna is used, oriented in the driving direction, towards the overtaking vehicle. The measurement starts as the car breaks the first light barrier. The second barrier is placed 3\,m after the first one for driving speed estimation. The distance between the transmitting and receiving vehicle is 15\,m, leading to about 50\,ns time of flight for the \ac{los} channel component. \\

\begin{figure}[!ht]
    \includegraphics[width=0.48\textwidth,page=3, trim=1.7cm 11cm 11cm 10.8cm, clip]{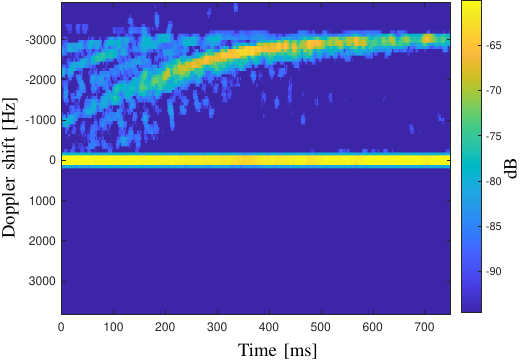}
		\caption{Doppler Power Profile.}
		\label{fig:DoppPowPro_case1}
\end{figure}
	
\noindent In order to better understand the cause of the distribution of stationarity regions, we calculate the Doppler power profile of the given channel, given for each \ac{lsf} instance as
\begin{equation}
    \boldsymbol{\hat{P}_{k_t,k_f}} = \dfrac{1}{M}  \boldsymbol{\hat{C}}_{\boldsymbol{k_t,k_f}} \mathbf{1}_M ,
\end{equation}
where $\mathbf{1}_M$ denotes a column vector of ones, with size $M$.
For this investigation, we set channel bandwidth to $B=272.7$\,MHz and \ac{lsf} time shift $\Delta_t=5$. Further, we used the \ac{lsf} size $N=100$, $M = 55$, for a potential stationarity time $N T_s=$12.9\,ms and bandwidth $M f_s=272.7$\,MHz, as shown in Fig.~\ref{fig:DoppPowPro_case1}. Here, we can distinguish between three time periods of the channel characteristics:
\begin{itemize}
    \item period \rom{1} - $t = [0,\dots,250]\,$ms: the overtaking vehicle is side-by-side with the receiver vehicle, multiple \acp{mpc} appear and cause a large Doppler spread due to the large reflecting surface of the overtaking \ac{suv}. Furthermore, as the \ac{suv} is moving, the relative speed to the receiver of the different parts of the car differ, causing \ac{mpc}s with different Doppler shift. Additionally, as the receive antenna is steered along the street, the \ac{suv} is outside the main lobe of the antenna's gain pattern, such that the \ac{los} component is dominant. 
    \item period \rom{2} - $t = [250,\dots,450]\,$ms: the  overtaking vehicle is inside the main lobe of the receive antenna gain pattern, causing a specular component comparable in strength to the LOS. Further, the \ac{suv} is oriented with a smaller surface to the receiver, leading to a smaller number of \ac{mpc}s.
    \item period \rom{3} - $t = [450,\dots,760]\,$ms: the vehicle is moving further away, leading to high attenuation of the reflected signal.
\end{itemize}

In this paper, we evaluate the size of stationarity regions in two scenarios. The first being \ac{los} case, where the direct path from the transmitter to the receiver dominates the channel. Further, we want to obtain a more common urban scenario, where the \ac{los} is blocked, and the wireless communication is performed only through the channel paths reflected of the other vehicles. Here, we describe the overtaking vehicle as a main source of reflecting components, moving with additional 15.8\,m/s relative to velocity of transmitter and receiver. We approximate this channel scenario by restricting the measurement data to Doppler shifts $\nu < -258$\,Hz, thereby simulating an \ac{nlos} scenario.

\section{Stationarity Criteria Definition}\label{sec_statCr}
In section \ref{Sec:lsf} we describe \ac{lsf} definition for the channels, where the statistical properties generally cannot be considered \ac{wss} in time, nor \ac{us} in frequency.
However, the channel is approximately \ac{wss}\ac{us} within a stationarity region.
In order to determine the regions of stationarity, first we calculate the \ac{lsf}, defined on the assumed local stationary region $T_{LSR} = N T_s$ in time and $B_{LSR} = M f_s$ in frequency. Second, we expand this region in time and frequency as long as the \ac{lsf} stays approximately constant. \\
As a metric defining whether we may extend the stationarity assumption over the neighboring \ac{lsf}s, we introduce a spectral distance metric called collinearity, as defined in \cite{b5}. We test the enlargement criteria independently in time and in frequency by introducing the collinearity metric for both domains. First, we define the metric to confirm the stationarity in frequency, expressed as
\begin{equation}
    \boldsymbol{\gamma^{(f)}}[k_f , k_{\Delta f}] = \dfrac{\sum_{k_t=1}^{K_t} \langle \boldsymbol{\hat{C}_{k_t,k_f}}, \boldsymbol{\hat{C}_{k_{ t},k_{\Delta f}}\rangle
    }_\mathrm{F} }{ \sqrt{ \sum_{k_t=1}^{K_t} \norm{ \boldsymbol{\hat{C}_{k_t,k_f}} }^2_\mathrm{F}
    \cdot 
     \sum_{k_t=1}^{K_t} \norm{ \boldsymbol{\hat{C}_{k_{t},k_{\Delta f}}} }^2_\mathrm{F} }},
\end{equation}
\noindent where $\langle \mathbf{A}, \mathbf{B} \rangle_\mathrm{F} =\sum_{i,j}\mathbf{A}[i,j] \mathbf{B}[i,j]$, for $\mathbf{A},\mathbf{B} \in\mathbb{R}^{I\times J}$, denotes Frobenius inner product, and $\lVert\mathbf{A}\rVert_\mathrm{F} = \sqrt{\langle \mathbf{A}, \mathbf{A} \rangle_\mathrm{F}}$ Frobenius norm. Further, $k_{\Delta f}$ is the frequency index of the shifted \ac{lsf}. As we are calculating the frequency region, over which the \ac{lsf} is approximately constant, we define a collinearity threshold $\gamma_{th} = 0.9$, above which we define the channel as stationary. Hence, we formulate the stationarity bandwidth as,
\begin{align}
    \boldsymbol{f_{stat}}[k_f] & = (M+(k_{\Delta f} -1) \Delta_f) f_s, \nonumber\\
    & \forall k_{\Delta f}  : \boldsymbol{\gamma^{(f)}}[k_f , k_{\Delta f}]>0.9.
\end{align}
In order to increase the delay resolution of \ac{lsf} before further examining the validity of the stationarity assumption, we update the size of local transfer function in frequency as $ M = \dfrac{\min\limits_{k_f}(\boldsymbol{f_{stat}}[k_f])}{f_s}$. Now we define the collinearity form for the stationarity investigation in the time domain
\begin{equation}
    \boldsymbol{\gamma^{(t)}}[k_t , k_{\Delta t}] = \dfrac{\sum_{k_f=1}^{K_f} \langle \boldsymbol{\hat{C}_{k_t,k_f}}, \boldsymbol{\hat{C}_{k_{\Delta t},k_{f}}\rangle
    }_\mathrm{F} }{ \sqrt{ \sum_{k_f=1}^{K_f} \norm{ \boldsymbol{\hat{C}_{k_t,k_f}} }^2_\mathrm{F} \cdot
     \sum_{k_f=1}^{K_f} \norm{ \boldsymbol{\hat{C}_{k_{\Delta t},k_f}} }^2_\mathrm{F} }},
\end{equation}
where $k_{\Delta t}$ denotes the time index of the shifted \ac{lsf}.

\noindent Finally, we express the stationarity time
\begin{align}
    \boldsymbol{t_{stat}}[k_t] & = (N+(k_{\Delta t} -1) \Delta_t) t_s, \nonumber \\
    & \forall k_{\Delta f}  : \boldsymbol{\gamma^{(t)}}[k_t , k_{\Delta t}]>0.9.
\end{align}

\section{Estimation of stationarity Regions} \label{sec_StatReg}
In this section, we present the estimation of stationarity regions in measured 60\,GHz channels. In order to reduce the variance, we apply a multitaper estimator with \ac{dpss} tapers as described in section \ref{Sec:lsf}. We define $IJ=4$, the total number of windows, by using two \ac{dpss} tapers in both time and frequency. Further, for bandwidth definition we set $N W_t = \tfrac2{T_s}$ and $M W_f = \tfrac{2.5}{f_s}$ in time and frequency, respectively. Additionally, to suppress the influence of noise on stationarity estimation, we set a noise threshold 10\,dB above the noise level. \\
A transfer function is approximately constant within the coherence time $T_c = \tfrac1{\overline{\nu}_H}$ and coherence frequency $f_c = \tfrac1{\overline{\tau}_H}$, as defined in \cite{b2}, where $\overline{\tau}_H$ is the \ac{rms} delay spread and $\overline{\nu}_H$ \ac{rms} Doppler spread. Therefore, it gives us a lower bound for the scaling of the local stationarity region. Our measured channel is described by maximum \ac{rms} delay spread $\overline{\tau}_{H, \text{max}} = 3.75\,$ms and \ac{rms} Doppler spread $\overline{\nu}_{H,\text{max}} = 250$\,Hz \cite{b7}. Hence, the minimal coherence time, valid over the whole measurement duration, is $T_{c,\text{min}}=4$\,ms, and minimal coherence frequency $f_{c,\text{min}}=266.7$\,MHz. \\
Further, we define a local stationarity region, laying inside the coherence region, spanning over $N=30$, $M=30$ samples in time and frequency, in which we approximate \ac{wss}\ac{us} condition as satisfied. This translates to $3.87$\,ms time period and $148.76$\,MHz bandwidth. Furthermore, for the time and frequency shift of consecutive \ac{lsf} regions, we set $\Delta_t = 5 $ and $\Delta_f = 5 $.\\ 
\begin{figure*}[b!t]
\begin{subfigure}{.5\textwidth}
    \includegraphics[width=0.98\textwidth,page=5, trim=1.7cm 19.74cm 10.8cm 1.7cm, clip]{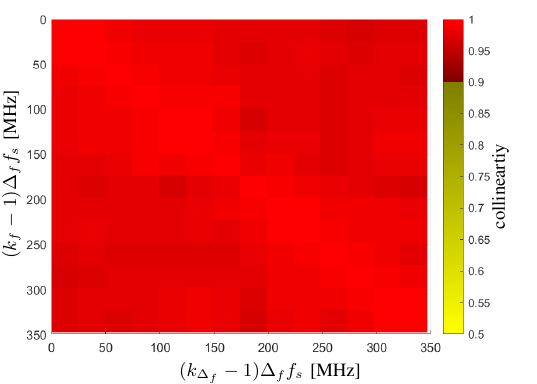}
    \caption{Stationarity bandwidth estimation}
	\label{fig:stat_f_los}
\end{subfigure}%
\begin{subfigure}{.5\textwidth}
\includegraphics[width=0.98\textwidth,page=5, trim=10.8cm 19.74cm 1.7cm 1.7cm, clip]{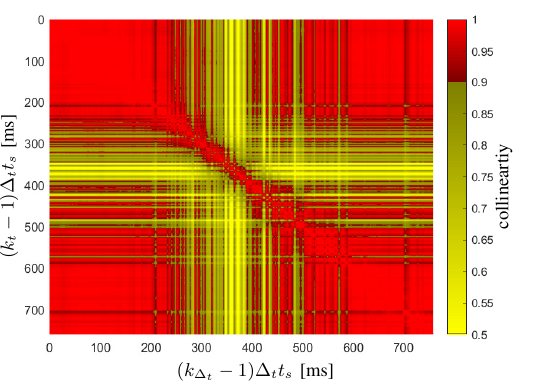}
  \caption{Stationarity time estimation}
		\label{fig:stat_t_los}
  
\end{subfigure}
\caption{\ac{wss}\ac{us} regions for \ac{los} scenario.}
\label{fig:wssus_los}
\end{figure*}
\begin{figure*}[ht]
\begin{subfigure}{.5\textwidth}
\includegraphics[width=0.98\textwidth,page=5, trim=1.7cm 11.8cm 10.8cm 9.64cm, clip]{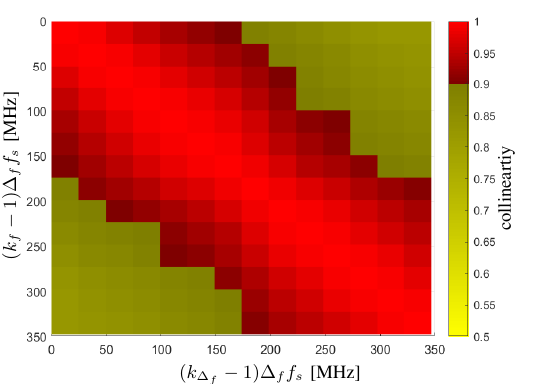}
		\caption{Stationarity bandwidth estimation}
		\label{fig:stat_f_nlos}
  
\end{subfigure}%
\begin{subfigure}{.5\textwidth}
\includegraphics[width=0.98\textwidth,page=5, trim=10.8cm 11.8cm 1.7cm 9.64cm, clip]{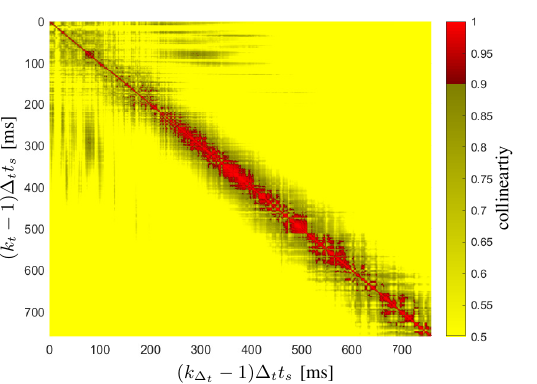}

		\caption{Stationarity time estimation}
		\label{fig:stat_t_nlos}
  
\end{subfigure}
\caption{\ac{wss}\ac{us} regions for \ac{nlos} scenario.}
\label{fig:wssus_nlos}
\end{figure*}
First, we investigate, the stationarity characteristics of the channel with a \ac{los} connection. Fig. \ref{fig:stat_f_los} demonstrates the channel stationarity over the whole bandwidth. We can notice that the collinearity values on the diagonal are equal to one, as the \ac{lsf} is compared with itself. As we move away from the diagonal, the value drops, but never under the defined limit of $0.9$. Therefore, for the investigation of stationarity in time, we increase the local stationarity region to span over the whole bandwidth, $M=103$. Now, we observe the channel stationarity in the time domain, Fig. \ref{fig:stat_t_los}. We can notice that the channel is stationary during the whole time period \rom{1} and \rom{3} defined in section \ref{sec_measScen}, as a \ac{los} connection between the fixed transmitter and receiver dominates the channel. Nevertheless, in the period \rom{2} when the reflected path is comparable in strength to the \ac{los}, we observe the mean stationarity time of $19.7$\,ms.\\
In order to further investigate the behavior of \ac{v2v} channels, we obtain the \ac{nlos} scenario, observing the channel components characterized by a Doppler shift $\nu < -258$\,Hz. Fig. \ref{fig:stat_f_nlos} shows the values of the collinearity function, over the increasing frequency shift between the two \ac{lsf}s. We observe that the channel is stationary at least over the bandwidth of $270$\,MHz. Furthermore, we enlarge the local stationary region to span over the observed stationary bandwidth, to M=55. Now we analyze the stationarity time, Fig. \ref{fig:stat_t_nlos}. During the period \rom{1}, the stationarity time is low, in the order of 5\,ms. The cause is the channel described by high number of \ac{mpc}s with variable delay and Doppler shift. Further, in the period \rom{2} we notice a quasi constant stationary time, with $16$\,ms average duration. During this period, the channel is dominated by one strong component, originating from a reflection of a vehicle moving with a constant relative speed of $15.8$\,m/s. 
In the period \rom{3} the stationarity time is rapidly changing as the strength of the specular channel component decreases and therefore, the residual noise plays higher role in the stationarity estimation.
\section{Conclusion} \label{sec_Conclusion}

We estimate stationarity regions in measured non-wide-sense stationarity uncorrelated scattering (non-\ac{wss}\ac{us}) millimeter wave (\ac{mmWave}) vehicle-to-vehicle (\ac{v2v}) channels. Here, we define a stationarity region, as a time-frequency area, over which the local scattering function (\ac{lsf}) is approximately constant, and therefore the \ac{wss}\ac{us} assumption is valid  approximately. As statistical quantifier, we use collinearity for independent analysis of stationarity in the time and frequency domains. 

\noindent
We conclude that the \ac{v2v} channels, dominated by a line-of-sight (\ac{los}) connection between two vehicles, which are moving in parallel, exhibit large stationarity regions. Channels with significant specular components originating from an overtaking vehicle moving at urban speed (relative to the others) feature a stationarity time of approximately $20$\,ms. 

\noindent
Furthermore, we analyze a case where the \ac{los} is blocked, and specular reflections can be dominant paths, which we consider as non-LOS (\ac{nlos}) scenario. The position of an adjacent vehicle relative to two communicating vehicles influences the size and shape of the stationarity region.
We observe that in the case of \ac{nlos} links with a large number of multipath components (MPCs),  the stationarity time becomes very short, in the order of just $5$\,ms. 
In contrast, in cases where the \ac{nlos} communication is dominated by one strong specular path, we observe longer stationarity times in the order of $16$\,ms. 
Moreover, in the \ac{nlos} \ac{v2v} scenario, we observe a stationarity bandwidth of $270$\,MHz or even larger.

\end{document}